# Long live the scientists: Tracking the scientific fame of great minds in physics

Guoyan Wang[a], Guangyuan Hu[b], Chuanfeng Li[c], Li Tang*[d, 1]

a. Department of Science and Technology Communication and Policy, University of Science and Technology of China, 96 Jinzhai Road, Hefei, Anhui Province, China, 230026.

b. School of Public Economics and Public Administration, Shanghai University of Finance Economics, 777 Guoding Rd., Shanghai, China, 200433

c. Key Lab of Quantum Information, University of Science and Technology of China, 96 Jinzhai Road, Hefei, Anhui Province, China, 230026.

d. Corresponding author, School of International Relations and Public Affairs, Fudan University, 220 Handan Rd., Shanghai, China 200433, Tel: (+86) 15000465884

**Abstract**

This study utilizes global digitalized books and articles to examine the scientific fame of the most influential physicists. Our research reveals that the greatest minds are gone but not forgotten. Their scientific impacts on human history have persisted for centuries. We also find evidence in support of own-group fame preference, i.e., that the scientists have greater reputations in their home countries or among scholars sharing the same languages. We argue that, when applied appropriately, Google Books and Ngram Viewer can serve as promising tools for altmetrics, providing a more comprehensive picture of the impacts scholars and their achievements have made beyond academia.

*Some say that a man dies three times. The first time is when his heart stops beating and he dies physically. The second is when people come to his funeral and his identity is erased from society. The third time is when nobody on the earth remembers him anymore. Then he is really dead.*

- "Dragon Raja" by Lee Yeongdo

## 1. Introduction

Books are the stepping stones to human progress. According to UNESCO (United Nations Educational, Scientific and Cultural Organization), the number of estimated published books in 2017 alone is up to 2.2 million.[2] Such a large collection is undoubtedly a rich archive of human history and civilization. Yet, as one of the most telling embodiments of knowledge stock and advancement, books have not captured sufficient attention in quantitative research evaluation.

Fortunately, with access to Google Books and the Google Books tool Ngram Viewer, scholars are now able to trace cultural evolution on a long time scale based on digitalized texts and trillions of words. This application of high-throughput data collection to study human culture can be traced back to Michel et al. (2011). In this pioneering study, the authors utilized the Google Books corpus and conducted text-based statistical analysis to trace cultural trends. That innovative research method soon captured academia's attention and was adopted in the arenas of digital history (Sternfeld, 2011), the history of science (Laubichler, Maienschein and Renn, 2013), economics (Roth, 2013), social psychology (Greenfield, 2013; Acerbi, 2013; Zeng and Greenfield, 2015), and cultural psychology (Pettit, 2016).

---

1 To whom correspondence should be addressed. Email: litang@fudan.edu.cn

2 Data source: http://www.worldometers.info/books/. Accessed on January 18, 2018.

Google Books has also been utilized to assess the fame of great scientists throughout history. Based on the word frequency of people's full names mentioned in books, Bohannon's Science Hall of Fame was built a as an objective evaluation of scientific fame over centuries (Bohannon, 2011a; 2011b). Moving beyond previous work, this paper utilizes both Google Books, which covers 36 million global digital books, and Google Scholar, which indexes 91 million academic items,[3] to examine the scientific fame of top physicists. We particularly focus on and compare two of the greatest physicists, Isaac Newton and Albert Einstein, depicting their fame evolution over centuries and exploring what they are famous for.

Our research reveals that the great minds are gone but not forgotten. Early scientists are still on the public's lips in modern society. Their scientific impacts on human history have persisted for centuries. This holds true for other prominent physicists as well. We also found that while Einstein's scientific fame has exceeded that of Newton among intellectuals since the mid-20$^{th}$ century worldwide, there is a different pattern of fame when the own-group preference is differentiated by the language of digitalized corpus. The computational analysis confirms that the influence of Einstein is largely related to his two contributions on general relativity and quantum theory, while the most frequently mentioned scientific achievements of Newton are the law of universal gravitation and the laws of motion.

This paper makes the following contributions to the literature. To begin with, this is the first attempt, within our best knowledge, to explore scientific fame based on the combination of both books and articles indexed by Google. In addition to depicting the evolution of fame of great minds, we also explore their most accredited achievements based on co-occurence analysis. Second, our study contributes to the discussion on the expected role of scientists in science promotion. By comparing the indicator for scientific impact (i.e., work being cited by scholarly publications) and the indicator for scientific fame (i.e., name being mentioned in books), our study sheds some light on how to gauge scientists' contributions beyond academia. We argue that, when applied appropriately, Google Books and Google Ngram Viewer can serve as promising tools for altmetrics, providing a more comprehensive picture of the impacts scholars and their achievements have made on society.

The rest of the paper is structured as follows. In the next section, we delineate our method and case selection justifications. Section 3 presents our analysis. In Section 4, following a summary of main findings, we conclude our paper discussing limitations and future research venues.

## 2. Method and Data

### 2.1 Notion and measurement

There is no agreed-upon definition of scientific fame or its measurement. The term can be traced back to the book *The Life of Sir Charles Linnæus*, a biography of a Swedish botanist, physician, and zoologist whose fame is centered on his enduring achievement of binomial nomenclature *(Stöver 1794).* Yet many scientists are ordinary folks and are little known to the public (Astin 1957; Menard 1971; Merton 1970). Some scholars have argued that the fame of scientists should be confined to professional achievements, while others believe scientific fame goes beyond academia (Menard 1971; Bohannon 2011a). Feist (2016) noted that regardless of either intrinsic or extrinsic research, the assessment of fame in science ultimately rests on productivity and its impact on advancing the research front. Previous studies often used being elected to prestigious societies or winning research awards or prizes as proxy indicators of scientific recognition or reputation (Bronk, 1976; Youtie, etc. 2013). Instead of relying on the subjective judgment of panel experts, Bohannon's Science Hall of Fame (2011a) innovatively uses the appearance of people's names in books to capture scientists' influence across different domains throughout history. This is the approach we adopt in tracking and recording the fame of great scientific minds and their achievements.

---

[3] The types of academic items Google Scholar includes are research articles, books, patents, case laws, and citations. In this research our search excludes patents and citations.

## 2.2 Case selection

The focus domain in this work is physics. Among the myriad scientists, we purposely choose Isaac Newton and Albert Einstein for illustration based on the following considerations.

To begin with, both Newton and Einstein are in the field of physics, which makes their comparison relatively free from discipline differences of publication distribution. And given that both names consist of two words with similar length, the quality of their retrievals is fairly comparative. As the two most influential physicists in the history of science (Whittaker, 1943; Baker, 1984), Newton and Einstein are appropriate candidates for evaluating the historical fame of individuals. Finally, who is more influential has been a topic attracting much attention in the global scientific community. The debate has not been settled for more than half a century (Gribbin, 1987; Graneau and Graneau, 1993). Successor scientists have commemorated them on special anniversaries, such as "Science 1943: Aristotle, Newton, Einstein, the three-hundredth anniversary of Newton's birth" and "New Scientist 1987: Newton vs. Einstein, the three-hundredth anniversary of Newton's theory of gravity." In 2005, the Year of World Physics and also the Centenary of Einstein's Special Relativity Theory, the UK Royal Society conducted polls of both academia and the public. The results showed that both Royal Society members and the British netizens surveyed considered Isaac Newton, a British scientist, to have greater influence on both science and humankind than Albert Einstein (*The Royal Society News*, 2005). It would be interesting to find out whether that also holds true beyond the geography of the UK.

## 2.3 Data

Our main datasets are Google digital books and scholarly articles. As shown in Table 1, the Google corpus used in this study consists of over 36 million global digital books and 91 million articles published since the 16$^{th}$ century.

(Table 1 insert here)

Table 1. Coverage of Google Books and Google Scholar

| Google corpus | Total returned hits | After 2000 | 20$^{th}$ century | 19$^{th}$ century | 18$^{th}$ century | 17$^{th}$ century | Before 1600 |
|---|---|---|---|---|---|---|---|
| Books | 36,540,000 | 14,600,000 | 14,700,000 | 5,680,000 | 808,000 | 319,000 | 433,000 |
| Scholar | 91,258,000 | 43,061,000 | 46,935,600 | 1,180,630 | 80,300 | 2 | 0 |

Note: This search was conducted through the library at London University in July 2015. For Google Books, the command we used is "allinurl:books" at https://books.google.com. For Google Scholar we set up the custom range of years in the advanced search. [4]

To gain a complete picture of these physicists' fame worldwide, we took a variety of languages into consideration. We first tested our search terms in 57 different languages using the Google Books search engine based on trials and errors.[5] The volumes and shares of the written languages of these books are displayed in Table 2. As shown, the dataset we analyzed covers more than 126 thousand books mentioning Newton and 149 thousand books mentioning Einstein since the 20th century. This finding seems opposite to that of the UK polls in 2005. Table 2 also reveals that, similar to other publication datasets, the coverage of Google Books is also biased in language (Lin 2012; Liu, et al. 2018; Liu et al., 2015). Books written in Chinese, Russian, Korean, and other non-English languages are highly underrepresented in the Google Books corpus. This caveat needs to be borne in mind when using this search engine for analysis.

(Table 2 insert here)

## 3. Analysis

### 3.1. Global fame over time

Figure 1 depicts the quantity of digitalized books and academic items mentioning Newton's and Einstein's full names. It needs to be pointed out that our returned hits consist of two scenarios: books mentioning Newton or Einstein and books written by either of them. But given the number of returned hits and the number of books even the most prolific scholar can write, it is reasonable to believe that the majority of the retrieved books are those mentioning them. Section 4 discusses this in more detail. Blue denotes Newton while red represents Einstein. The dotted lines mark the deceased years of the two great scientists, respectively. Please note that any key word search is only effective for accessible content. In our case, only Google Books with a preview available can be searched for names in the text.

(Figure 1 insert here)

In Figure 1A, the number in the upper left ($1e^4=10^4$) is the unit of the absolute number of books indexed in the Google Books search. For Newton, the most glorious period of his historical influence is between 1680 and 1880, when records mentioning Newton amount to more than 60% of all records mentioning him. As demonstrated, the deaths of these great scientists do not dim their glory. On the contrary, the popularity of both physicists has risen dramatically since the 1980s. One may wonder what happened in the period of 1880 to 1980, when relatively fewer books talked about Newton and Einstein. One speculation is that two dark clouds appeared in the world of physics at the end of the 19$^{th}$ century: the Michelson–Morley experiment and the ultraviolet catastrophe in Planck's law (Kelvin, 1901). They overturned the framework of the old theory system, leading to the birth of quantum and relativity theories (Hoover, 1977; Sanghera, 2011). This, when combined with the breakout and aftermath of World Wars I and II, partially explains the silence of Newton's voice from 1880 to 1980.

---

[4] It needs to be pointed out that depending on the search date and location, the returned hits vary. Take Google Books, for example. We repeated the same search in July 2018, and the returned hits are not 36,540,000 as in July 2015 in the UK but are 38,400,000 in China and 39,300,000 in South Korea. This pattern also holds for Google Scholar. As shown in Table 1, the coverage of Google Scholar was 43,061,000 articles published over the period of 2000–2015 when the search was conducted at London University in July 2015. In August 2018, its coverage of the same period extends to 46,840,000 and 48,720,000 articles when searched in China and South Korea, respectively. Our experiences echo previous findings on the coverage scope, growth rate, and indexing quality of Google Books and Google Scholar (Abdullah & Thelwall, 2014; Fagan, 2017; Halevi et al. 2017). For an extensive review of the pros and cons of Google Books and Google Scholar as well as their utilities in research assessment, please refer to Google Scholar Digest at http://googlescholardigest.blogspot.com/p/bibliography.html.

[5] The variations of name order (e.g., "Isaac Newton" and "Newton, Isaac") were taken into consideration.

Figure 1B maps the scientists' influences in academia via a Google Scholar search engine. The y values, i.e. relative counts, were computed by dividing the number of Google Scholar indexed articles mentioning the scientist in a given year by the total number of Google Scholar indexed articles in that year. As shown, Newton leads in fame until 1918, then both scientists parallel for 30 years. The year of 1948 is the watershed. Since then, Einstein has gained more popularity than Newton worldwide.

One key message conveyed by Figure 1 is that globally Einstein seems to enjoy greater fame than Newton, which is contrary to the results of the 2005 UK polls mentioned previously. If the findings of both the surveys and the computational analysis are valid, one possible explanation for this discrepancy is that scientific fame varies by location. People are more recognized within their own group (Egghe and Rousseau, 2004; Tang et al., 2015), be it by country or by shared language. To test

Table 2. Retrieved hits in Google Books by different languages

| Scientist | Search word | Retrieved hits in 57 languages | 2000–2015 (prop.) | 1900–1999 (prop.) |
|---|---|---|---|---|
| **Isaac Newton** (1643–1727) | Isaac Newton | Records in 49 languages with the same name spelling as in English | 95.78% | 98.18% |
| | Исаак Ньютон | Russian | 2.01% | 1.28% |
| | 아이작 뉴턴 | Korean | 1.17% | 0.12% |
| | アイザック・ニュートン | Japanese | 0.51% | 0.31% |
| | "艾萨克 牛顿"OR"艾薩克 牛頓" | Chinese (simplified or traditional) | 0.21% | 0.06% |
| | إسحاق نيوتن | Arabic | 0.18% | 0.04% |
| | אייזק ניוטון | Hebrew | 0.15% | 0.02% |
| | আইজাক নিউটন | Bengali | 0.01% | 0.00% |
| | आइज़ैक न्यूटन | Hindi | 0.00% | 0.00% |
| | **Total retrieved hits of Newton** | | **125593** | **36434** |
| **Albert Einstein** (1879–1955) | Albert Einstein | Records in 49 languages with the same name spelling as in English | 96.25% | 96.78% |
| | Альберт Эйнштейн | Russian | 2.35% | 2.08% |
| | אלברט איינשטיין | Hebrew | 0.42% | 0.62% |
| | アルバート・アインシュタイン | Japanese | 0.35% | 0.22% |
| | 알버트 아인슈타인 | Korean | 0.26% | 0.08% |
| | "阿尔伯特・爱因斯坦"OR"阿爾伯特・愛因斯坦" | Chinese (simplified or traditional) | 0.21% | 0.17% |
| | البرت اينشتاين | Arabic | 0.05% | 0.06% |
| | अल्बर्ट आइंस्टीन | Hindi | 0.01% | 0.00% |
| | আলবার্ট আইনস্টাইন | Bengali | 0.00% | 0.00% |
| | **Total retrieved hits of Einstein** | | **149169** | **38154** |

Data source: Google Books search. Data accessed in January 2016.
Note: In the other 49 languages such as German, French, Italian, Spanish, and Portuguese, both name spellings are exactly the same as in English.

this speculation, we utilized the language option of Google Books Ngram Viewer and examined the physicists' historical influence by various languages.[6]

---

[6] Since 2010 the Google Books Ngram Viewer has been available at http://books.google.com/ngrams. Its source, i.e. Google Books, was generated through Partner Program and the Library Project (please refer to https://www.google.com/intl/en/googlebooks/about/). These millions of books published between 1500 and 2008 were digitally scanned and the corpus was winnowed. For more details please see Michel et al. (2011) and Twenge et al. (2012). The searchable corpora with the Google Books Ngram Viewer were developed by the Google Team and can be

**Figure 1. Global fame of Newton and Einstein**

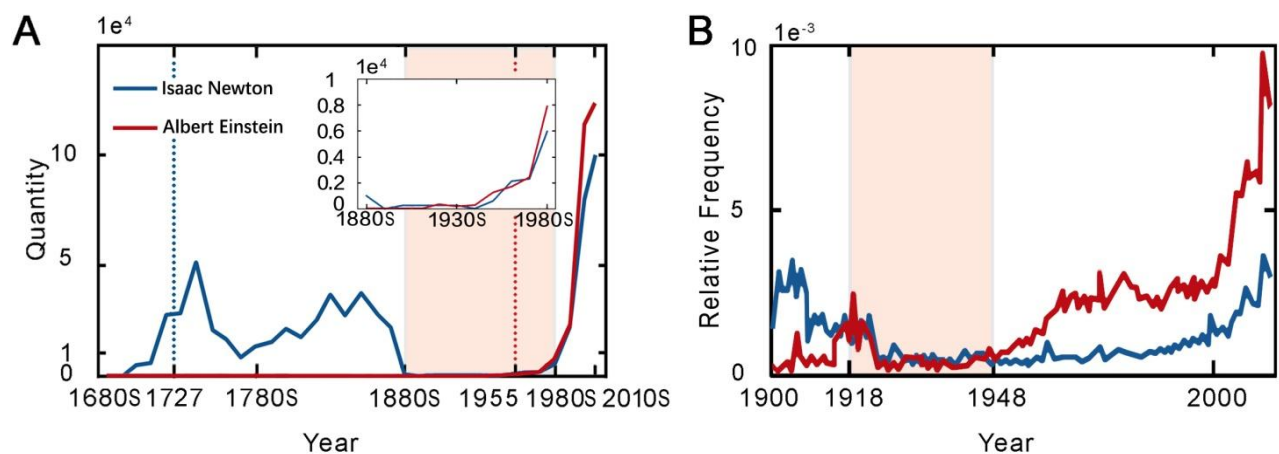

Data source: Panel A is Google Books search. The time coverage of the main graph is from 1680s to 2010s with an inset zooming in on the period of 1880s–1980s; Panel B is Google Scholar search. The time coverage is from 1900 to 2012. Accessed in January 2016.
Forty-nine languages with the same name spelling as in English are included.

### 3.2. Global fame moderated by language

Figure 2 was generated with Google Books Ngram Viewer with a default smoothing of three.[7] The four panels illustrate the scientific fame of Newton and Einstein in books written in British and American English together, British English alone, American English alone, and German, respectively. Both scientists seem to be preferred in their own groups. In the British English case, Newton has consistently enjoyed more popularity than Einstein throughout the last 100 years (Panel B). This finding is consistent with 2005 UK polls results. In sharp contrast, Einstein, a German-origin physicist who later immigrated to the US, has been mentioned far more than Newton since the mid or late 20th century in British and American English together, American English alone, and German books (Figures 2A, 2C, and 2D). This comparison of citing book language uncovers a high relevance of own-group preference, or approximately location of fame.

(Figure 2 insert here)

---

downloaded at http://storage.googleapis.com/books/ngrams/books/datasetsv2.html. Please note that Google Books Ngram Viewer cannot be used to determine how often a search term occurs in a book.

[7] The default smoothing parameter of Google Books Ngram Viewer is set at 3. Taking the year of 2000 as an example, a smoothing of 3 means that the data shown for 2000 are a mathematical average of original data for 2000 plus three values before and three after 2000 (i.e., a moving average of seven-year data, including 1997, 1998, 1999, 2000, 2001, 2002, and 2003). This setting avoids spikes and makes trends more apparent (Michel et al., 2011; Greenfield, 2013). More information on graphs generated by Google Books Ngram Viewer is available at https://books.google.com/ngrams/info.

# Figure 2. Scientific fame by language

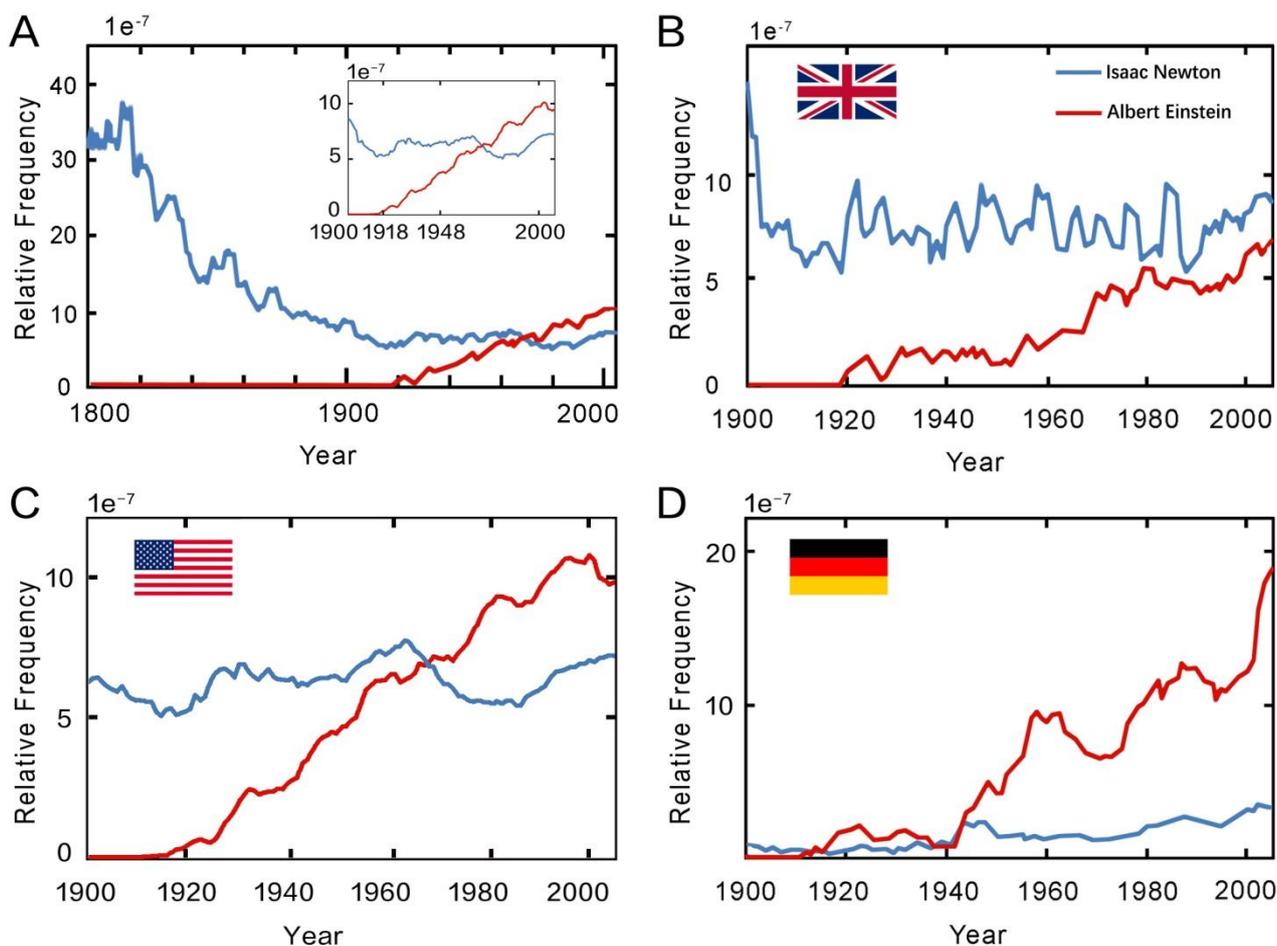

Data source: Google Books Ngram Viewer. Accessed in January 2016.
Panel A: Only books in English (both British English and American English) are considered.
Panel B: Only books in British English are considered.
Panel C: Only books in American English are considered.
Panel D: Only books in German are considered.
The time coverage is from 1800 to 2008 for Panel A and from 1900 to 2008 for Panels B~D.

## 3.3 Achievements they are famous for

Scientists are commemorated for various reasons. Hawking and Israel (1989) note that the greatest contributions made by Newton in the history of science are the law of universal gravitation and the three laws of motion. In addition to physics, Newton also contributed to the fields of mathematics, astronomy, and the philosophy of nature (Newton, Cohen, & Schofield, 1959; Newton, 1999). In comparison, the most notable achievements of Einstein are quantum mechanics and general relativity (Whittaker, 1943; Torre, et al., 2000). Then, a question arises: what are they famous for as evidenced by the Google corpus?

As demonstrated in Figure 3A, the contributions of Newton surpass the Laws of Motion and the Law of Gravity. This is particularly true prior to mid-1800. As shown the name of Isaac Newton is mentioned in Google Books far more than his most accredited physics theories are.

(Figure 3 insert here)

**Figure 3. The major scientific contributions of Newton and Einstein**

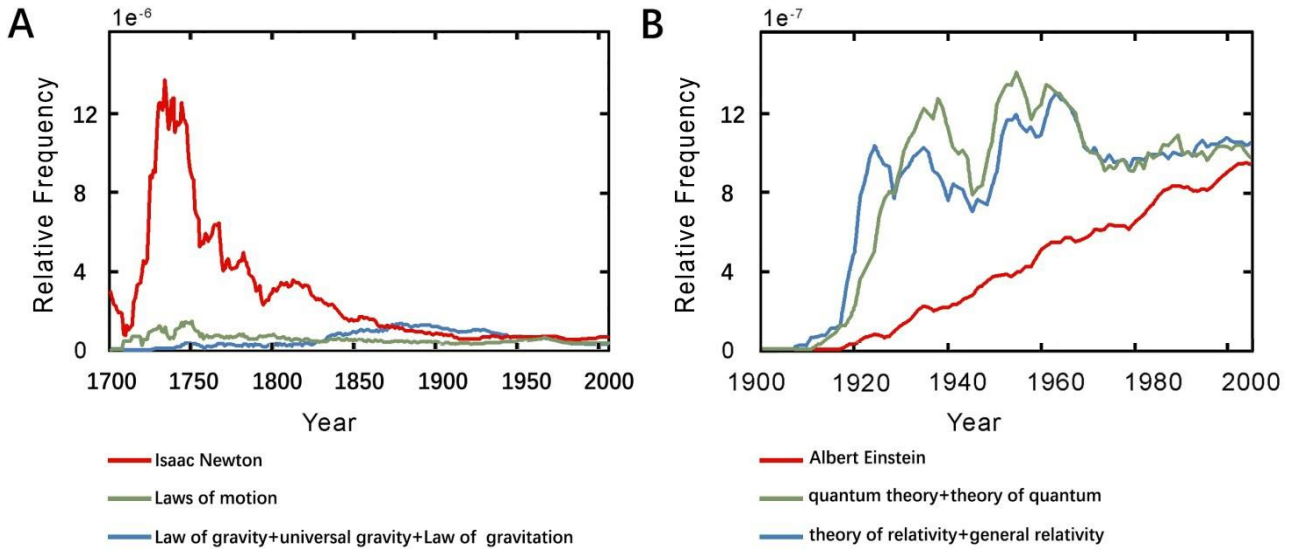

Data source: Google Books Ngram Viewer. Accessed in December 2017.

The time coverage is from 1700 to 2008 for Panel A and from 1900 to 2008 for Panel B.

One may notice the close intertwining or convergence of lines representing scientific achievements and the scientists' names in later periods. But the approximately same frequency of mentioning the achievements and the scientists' names does not mean that the mentions appear simultaneously within the same corpus books. So, we next conducted co-occurrence analysis to examine to what extent these scientists' most accredited achievements contribute to their enduring reputations (Table 3).

(Table 3 insert here)

As depicted in Table 3, books mentioning Isaac Newton often deal with the law of universal gravity (15%) and calculus (8%). The influence of Albert Einstein is highly correlated to the theory of relativity (about 28%) and quantum theory (almost 17%). This, on one hand, reveals what achievements contribute to the glory of Newton and Einstein; on the other hand, the residue (i.e., the uncorrelated proportion) suggests other elements (such as other achievements or even anecdotes) exist.[8]

---

[8] Our manual examination of the remaining books mentioning Newton but not gravity, gravitation, calculus, or motion suggests that many are related to natural philosophy, alchemy, and theology. In addition to mentions of relativity, quantum, photoelectric and mass-energy, books mentioning Einstein also discuss his ethical and philosophical views on cosmic religion, his educational philosophy, and even his private life.

Table 3. Co-occurrence analysis of contributions and physicists

| Key words | 1900–2015 | |
|---|---|---|
| | Records in Books | % of Total |
| Isaac Newton | 156,100 | / |
| Newton and gravity (including law of gravity, universal gravity, universal gravitation, law of gravitation) | 23,700 | 15.2% |
| Newton and calculus | 12,880 | 7.9% |
| Newton and laws of motion | 10,440 | 6.7% |
| Albert Einstein | 178,200 | / |
| Einstein and relativity (including general relativity and special relativity) | 50,130 | 28.1% |
| Einstein and quantum | 30,020 | 16.9% |
| Einstein and photoelectric | 4,200 | 2.4% |
| Einstein and mass-energy | 2,420 | 1.4% |

Data source: Google Books search. Data accessed in January 2016.

### 3.4 Other great minds in physics

We extended our analysis to other great physicists in history. We first compiled a list of 234 of the most influential scientists based on the following four books:

*The 100 Most Influential Scientists of All Time* (Rodgers 2009)
*The 100 Most Influential Scientists* (Simmons 2009)
*100 Scientists Who Changed the World* (Tiner 2002)
*The Most Influential Scientists in Human Civilization* (Luo 2012)

After removing scientists born before the year 1 BC, such as Aristotle and Plato, we next searched each name in Wikipedia and allocated the scientists to different research fields. We then retrieved their name appearance frequency in all Google Books published after 2000. The top 20 physicists listed in Table 4 show that in the field of physics, Albert Einstein, Max Planck, Isaac Newton, Blaise Pascal, and Galileo Galilei are the top five most influential and best-known physicists in the 21st century.

(Table 4 insert here)

Table 4 lends further support that great scientists are gone but not forgotten.[9] As shown, the early scientists such as Newton, Pascal, and Galilei are still on the public's lips in contemporary society. Undoubtedly, the achievements of these scientific pioneers have become the cornerstones of human scientific knowledge. But it is also true that besides scientists' contributions to scientific advancement, their scientific spirit, anecdotes, and other special attributes may remain as well. Students and the public learn about the great scientists via scholarly publications, textbooks, and popular science books. When a scientist becomes a public figure, his life experiences and achievements are often adapted into various biographies and even enlightening stories. In stories such as "Newton and His Apple," the scientific spirit has been used to encourage and to inspire each succeeding generation for hundreds, if not thousands, of years.

---

[9] The scientific fame of these 20 physicists persists after they pass away.

Table 4. Top 20 physicists ranked by scientific fame

| Rank | Name of Physicist | Nationality of Origin | Year of Birth | Rec. of being mentioned | Rank | Name of Physicist | Nationality of Origin | Year of Birth | Rec. of being mentioned |
|---|---|---|---|---|---|---|---|---|---|
| 1 | Albert Einstein | German | 1879 | 145,000 | 11 | Michael Faraday | UK | 1791 | 49,500 |
| 2 | Max Planck | German | 1858 | 129,000 | 12 | James Watt | UK | 1736 | 47,900 |
| 3 | Isaac Newton | UK | 1643 | 121,000 | 13 | Clerk Maxwell | UK | 1831 | 43,100 |
| 4 | Blaise Pascal | France | 1623 | 118,000 | 14 | Enrico Fermi | Italy | 1901 | 41,400 |
| 5 | Galileo Galilei | Italy | 1564 | 96,100 | 15 | William Thomson | UK | 1824 | 40,700 |
| 6 | Stephen Hawking | UK | 1942 | 82,800 | 16 | Robert Hooke | UK | 1635 | 40,200 |
| 7 | Niels Bohr | Denmark | 1885 | 80,100 | 17 | Max Born | German | 1882 | 38,000 |
| 8 | Joseph Henry | USA | 1797 | 79,100 | 18 | Wolfgang Pauli | Austria | 1900 | 37,200 |
| 9 | Richard Feynman | German | 1918 | 65,300 | 19 | Robert Oppenheimer | USA | 1904 | 36,400 |
| 10 | Werner Heisenberg | German | 1901 | 55,400 | 20 | Thomas Young | UK | 1773 | 36,400 |

Data source: Google Books search. Data accessed in January 2016.
Time coverage: 2000–2015

This also holds for Stephen Hawking. In the list of top 20 physicists in the 21$^{st}$ century shown in Table 4, Hawking ranks sixth and far ahead of some classic scientists including James Watt, Michael Faraday, and so on. In addition to Hawking's great scientific achievements, his passion in promoting science given his special physical condition has been seen as inspiring to the public and especially younger generations.

## 4. Conclusion and discussion

### 4.1 Summary

Big data is an important theme in natural and social sciences today. Adopting and expanding the approaches developed by Bohannon (2011a, 2011b), Michel et al. (2011), and Lin et al. (2012), this work exemplifies how to track the historical achievements and influence of great scientific minds. Utilizing the language option of Google Books Ngram Viewer, we find evidence in support of own-group preference, which explains the discrepancy of the UK survey and computational analysis results based on the Google corpus. The co-occurrence analysis demonstrates the main contributions that scientists are known for by later generations

The wheel of history is always moving forward, with new developments emerging and old theories gradually fading away. This is an inexorable law of history, with each scientist's face and influence having a certain vitality cycle. Yet our study suggests the great minds live long intellectually. A scientist can only live several decades in physical form, but his contributions can live on in human history for hundreds or even thousands of years. Today when we talk about pioneers of science before the Christian era such as Aristotle, Euclid, and Archimedes, we find that these names have been written into history, respected and eulogized repeatedly by generations of newcomers. This may be the highest level of the meaning of life that scientists are pursuing.

## 4.2 Future research

It is an accepted practice to use citations to measure the scientific impact of researchers or their research output. Yet for an academic paper, the citation life cycle is usually about 3 to 5 years, and the count of citations often declines drastically afterwards (Glänzel and Schoepflin 1995; Tang 2013; Rogers 2010). For a long time bibiometricians have used or abused this impact factor while also acknowledging the difficulties in quantifying long-term scientific impact (Wang, Song and Barabási, 2013; Waltman 2016). With the development of Google digitalized texts and computational analytical methods, we can track names being mentioned and assess the evolution of scientists' fame on a global scale over centuries. Given the variety of book types indexed in Google, this opens another possible research venue of altmetrics: measuring the influence of scientists beyond academia (Costas et al., 2015).

In his critical overview of altmetrics, Bornmann (2014) pointed out that no accepted framework existed to measure societal impact. Different from the existing altmetrics, which assess the impact of scientific work mostly on social media, we argue that when applied appropriately, Google Books can serve as a promising source of data for altmetrics, and the appearance frequency of scientists' names in books can serve as alternative metrics, which are still in great flux, to capture scientists' intellectual contributions to a broader public (Bornmann, 2014; Fenner, 2013; Piwowar, 2013). In the future it would be interesting to differentiate the fame of a scientist within academia and among the public with the categories provided in Google Books search (textbook, fiction, scholarly book, etc.). This paper focuses on physics; research looking into scientific fame of figures in other disciplines or even across disciplines is also worthy of further examination.

## 4.3 Limitations

In spite of their potential value for evaluation research, we would like to raise several notes of caution regarding the data and the method.

There are two sources of false positives. Searching individuals' names in books or scholarly publications can yield two types of results: works written by them and their achievements cited or mentioned by others. Given that Google Books and Ngram Viewer cannot differentiate the specific locations of search terms within books, false positives are inevitable. The good news is that even for an extremely prolific scientist, the actual number of articles one can write and publish is only in the hundreds. As an illustration of authorship versus citation, we searched Alexandre Dumas (1802–1870) who, though not a scientist, was a prolific French writer who dedicated his life to writing and published 357 books. We searched his full name in Google Books on December 20, 2017, and retrieved 23,400 hits. This suggests that the number of times mentioned can serve as a proxy indicator of fame of prominent individuals and is not a reflection of their authorship.

The other source of false positives is name disambiguation. For instance, searches for just "Newton" could result in mentions of other people or even the International System of Units (SI) unit of force. The common name problem, translation, and transliteration can also lead to an overestimation of scientists' fame. On the other hand, name changes over time associated with marriage and other reasons can produce false negatives, or an underestimation of reputation. In this sense the measurement of scientific fame works best for great scientists or scientists with unique names. For more details, please refer to MacRoberts and MacRoberts (1989) and Tang and Walsh (2010).

We also need to be aware of the issues of eponymy and obliteration, which have been discussed rather comprehensively by McCain (2011). When a scientist's achievements are so important as to become common knowledge, it is possible that only the achievements themselves rather than the scientist's full name are referenced. For instance, when people talk about the law of gravity, they may not mention Newton's name at all. In this case, using full name as a proxy causes an underestimation of fame. As McCain noted, "it is often observed that very well-established concepts, experimental methods, empirical findings, and models are mentioned in the literature without being linked to their originators" (p. 1413). This in our case will incur false negatives of recording scientists' fame. However, we argue that a scientist's full name never being mentioned once throughout a whole book is rare and thus has minimal impact on our findings.

It should be noted that for any computational analysis the amount of retrieved data should be sufficiently large. Scientific fame is no exception. Despite its potential, researchers should be aware of the caveats of the Google corpus, such as language bias against non-English languages, the quality heterogeneity of indexed books, and whether it is appropriate only for highly influential scientists rather than general scientists, when conducting fame analysis. It needs to be pointed out that though Google Books and Ngram Viewer make it possible for us to analyze already famous scientists' fame, it would not be possible to identify those who are not yet famous scientists.[10]

Another caveat of this novel measurement is that it may not separate mentions of a famous figure for infamy rather than mentions of the figure for positive glory. Fortunately, in our case of two well-acknowledged great minds in science, the majority of the books on Newton and Einstein are based on a positive or at least a neutral evaluation. This saves us from the concern about evaluating controversial figures.

### Acknowledgements

The research is supported by the National Social Science Foundation of China (#14CXW011), Korea Foundation for Advanced Studies (International Scholar Exchange Fellowship 2017-2018), and the Humanities and Social Sciences Project of the Ministry of Education of China (#WKH3056005).We would like to express our sincere gratitude to two anonymous reviewers and Professor Ludo Waltman for their insightful suggestions which substantially improved this manuscript. We thank Professor Fei Xu for the inspiring discussion, Jiafei Shen and Feifei Han for their help in handling the figures. Author contributions: G.W., C.L., and L.T. designed research; G.W. analyzed data; and L.T and G.H. wrote the paper. The authors declare no conflict of interest.

---

[10] We would like to thank one anonymous reviewer for directing us to this point.